# Ensemble-based characterization of unbound and bound states on protein energy landscape

**Anatoly M. Ruvinsky[1*], Tatsiana Kirys[1,2], Alexander V. Tuzikov[2], and Ilya A. Vakser[1,3]**

[1]Center for Bioinformatics, The University of Kansas, Lawrence, Kansas 66047, USA

[2]United Institute of Informatics Problems, National Academy of Sciences, 220012 Minsk, Belarus

[3]Department of Molecular Biosciences, The University of Kansas, Lawrence, Kansas 66045, USA

[*]To whom correspondence should be addressed: email: ruvinsky@ku.edu

**Abstract**

Characterization of protein energy landscape and conformational ensembles is important for understanding mechanisms of protein folding and function. We studied ensembles of bound and unbound conformations of six proteins to explore their binding mechanisms and characterize the energy landscapes in implicit solvent. First, results show that bound and unbound spectra often significantly overlap. Moreover, the larger the overlap the smaller the RMSD between bound and unbound conformational ensembles. Second, the analysis of the unbound-to-bound changes points to conformational selection as the binding mechanism for four of the proteins. Third, the center of the unbound spectrum has a higher energy than the center of the corresponding bound spectrum of the dimeric and multimeric states for most of the proteins. This suggests that the unbound states often have larger entropy than the bound states considered outside of the complex. Fourth, the exhaustively long minimization, making small intra-rotamer adjustments, dramatically reduces the distance between the centers of the bound and unbound spectra as well as the spectra extent. It condenses unbound and bound energy levels into a thin layer at the bottom of the energy landscape with the energy spacing that varies between 0.8-4.6 and 3.5-10.5 kcal/mol for the unbound and bound states correspondingly. At the same time, the energy gap between the two lowest states in the full ensemble varies between 0.9 and 12.1 kcal/mol. Finally, the analysis of protein energy fluctuations showed that protein vibrations itself can excite the inter-state transitions, thus supporting the conformational selection theory.

# Introduction

Relationships between protein energy landscape, structure, and function have been a subject of numerous studies resulted in the development of the funnel shape energy landscape theory (1-5). This theory has been further extended by the conformational selection paradigm to include the ensemble-based description of proteins and protein-protein interactions (5-8). The concept suggests that bound and bound-like conformations may co-exist in solution within a large ensemble of unbound conformations. By shifting equilibrium in the unbound ensemble towards the bound-like conformations, binding forces select a bound conformation corresponding to the free energy minimum. Recent studies focused on reconstruction of the native ensembles (9-12). An ensemble of ubiquitin structures reflecting dynamics up to the microsecond time scale was refined against residual dipolar couplings. All crystallographically determined bound conformations of ubiquitin were found within 0.8 Å root mean square deviation (RMSD) of the $C_\alpha$ atoms (11). Another RDC-optimized ensemble of ubiquitin consistent with the microsecond time scale dynamics was created by Monte Carlo sampling of the ''Backrub'' motions (12). Cold denaturation and protein encapsulation were combined with NMR to probe the ensemble (13). Single-molecule experiments corroborated the theory of multiple interconverting conformations and revealed their relation to the fluctuating catalytic reactivity (14). Room-temperature X-ray crystallography was able to detect such conformations in proline isomerase (15). Best *et al.* (9) showed that an ensemble of highly homologous X-ray structures can also reproduce structural diversity in the native ensemble probed by NMR spectroscopy in solution. A protocol combining molecular dynamics (MD) simulations of an X-ray structure with information from the NMR relaxation experiments has been suggested for studying protein conformational ensembles in solution(16). In general MD simulations have been instrumental in mapping the conformational space (17-19). Alternative methods for generating conformational ensembles without solving explicit equations of motion have been actively developed (see Ref. (20) for a review). Large conformational ensembles are routinely used in protein structure prediction (21, 22) and studies of allosteric interactions (23, 24).

Despite the significant progress achieved in generating protein ensembles, their energetic properties and relation to the unbound-to-bound conformational changes are not well understood (25). How to generate a bound-like structure from the unbound one is one of the main problems in structure prediction of protein complexes. Although MD simulations showed that the interface side chains – "anchor residues" - sample bound-like conformations (26, 27), criteria for selecting such conformations from the MD snapshots are yet to be determined. Current docking protocols are much more successful when bound conformations are used, but become less reliable in a common case when only unbound structures are known (28, 29). To advance the docking protocols, the relation between the energy landscape and conformational changes upon binding should be unraveled. Recent large-scale studies of conformational changes upon binding focused on the relationship between single bound and single unbound conformations (30-32). However, how well the change between two selected conformations characterizes transition between the unbound and bound states within conformational ensembles as well as the transformation of the (free) energy landscapes is still unclear.

In this study, we investigate structural similarity between the ensembles of bound and unbound conformations for six proteins and characterize their energy landscapes in implicit solvent. We consider impact of the energy minimization in the Generalized Born (GB) model on the distance between the ensembles of bound and unbound conformations and the ensembles' spectral properties (the energy spacing, the spectrum gap between the lowest states, and the spectrum width). Our focus on the energy minimization in implicit solvent was motivated by a recent study (33) showing that ranking protein structures by minimized GB energies can distinguish the near-native structures from decoys better than ranking based on the energy minimization either in vacuum or explicit solvent. First, our study shows that although the shortly minimized GB energies of the bound and unbound ensembles often significantly overlap, the center of the unbound spectrum tends to have a higher energy than the centers of the bound spectrum of the dimeric and multimeric states. Moreover, the larger the overlap the smaller the RMSD between bound and unbound conformational ensembles. Second, the existence of the structurally different equipotential states in both ensembles suggests that unbound states have larger entropy than the bound states. The entropy-driven modeling of the unbound-to-bound conformational changes suggests a novel direction in advancing protein-protein docking algorithms, which, in fact, commonly neglect entropy effects. Third, the results show that the bound conformations of the RNase A interface pre-exist in the unbound ensemble, indicating conformational selection as the binding mechanism. Pancreatic trypsin inhibitor, ubiquitin, and lysozyme C also have high similarity between bound and unbound interfaces as well as small deviations that can be attributed to flash cooling (34, 35) or variations in the crystallization conditions. Fourth, the exhaustively long minimization by the Adopted Basis Newton-Raphson algorithm results in small mostly intra-rotamer adjustments that dramatically reduce the distance between the centers of the bound and unbound spectra as well as the spectra extent. It condenses unbound and bound energy levels into a thin layer at the bottom of the energy landscape. At the same time, the whole spectrum from the shortly minimized states to the bottom of the folding funnel can cover up to 40.3% of the lowest energy, indicating that the folded states may significantly differ in energy. The average energy spacing at the bottom of the energy landscape varies between 0.8-4.6 and 3.5-10.5 kcal/mol for the unbound and bound states correspondingly. The energy gap between the two lowest states varies between 0.9 and 12.1 kcal/mol. Finally, the results show that protein vibrations itself can stimulate the inter-state transitions, thus supporting the conformational selection theory. We suggest an approach for estimating the number of normal modes involved in conformational transition and show that, on average, 20 low-energy normal modes are needed to describe transition between two neighboring energy states. At the same time, transitions between the two lowest states may involve an order of magnitude larger number of the modes.

## Results and Discussion
*Bound and unbound energy bands*

Figure 1 shows minimized energy spectra of six proteins (see Methods) represented by the ensembles of their conformations determined by X-ray crystallography and NMR (Tables 1 and S1). Short and long minimizations (SM and LM) were applied to characterize the topography of the energy landscape in the Generalized Born model (see Methods). The energy minimizations caused small in-rotamer re-adjustments of the exposed side chains resulting in a typical RMSD≤0.7 Å between the minimized and the non-minimized structures, which did not substantially change neither the sizes of the bound and unbound ensembles nor the distance between them (Table 2). Nevertheless, these changes were enough to significantly condense both spectra of the unbound and bound proteins. Figure 1 shows that the span of the spectra and the spacing between energy minima after LM are significantly smaller than that after SM. The ratio between the overall energy span (including the SM and LM bands) in the unbound ensemble and the lowest energy in the protein spectrum is 40.3% for ovomucoid, 13.7% for PTI, 26.9% for ubiquitin, 21.5% for RNase A, 22.8% for CheY and 24.2% for lysozyme C. Excluding the lowest ratio for PTI as an outlier, the ratio for other five proteins decreases 1.7 times, with a 2.5 times increase of the number of atoms from ovomucoid to lysozyme C. The ratio decrease is expected because the lowest energy is a function of the total number of protein residues, whereas the energy extent relates mainly to the surface residues that are able to change their conformations in solution. The outlying ratio for PTI may result from the insufficient size of its unbound ensemble, which is the smallest among the proteins in our set (Table 1). Ovomucoid has 51 residues, compared to 56 residues of PTI, but its unbound ensemble is 4.6 times larger than the ensemble of PTI.

The ratio between the energy span of the LM band and the lowest energy is 7.0% for ovomucoid, 4.9% for PTI, 14.0% for ubiquitin, 5.5% for RNase A, 5.0% for CheY and 3.1% for lysozyme C (Fig. 2). Thus, LM reduces the width of the energy bands, condensing protein states into a thin layer at the bottom of the energy landscape. The energy distance (the ruggedness) between the centers of the SM and LM energy bands was calculated as an arithmetic mean of energies in a protein ensemble after short and long minimizations (see Methods). Interestingly, the energies of the unbound ensembles decrease more upon minimization than the energies of the bound ensembles (Fig. S1) despite the fact that both ensembles have equipotential energy levels (Fig. 1). On average, the unbound proteins lose 0.6 kcal/mol per heavy atom or 4.6 kcal/mol per residue. The bound structures from dimers and multimers lose less: 0.4 and 0.3 kcal/mol per heavy atom or 3.1 and 2.5 kcal/mol per residue accordingly.

The centers of the unbound energies after SM were higher than the centers of the bound energies for all proteins, with the exception of ubiquitin. This suggests that the unbound-to-bound conformational changes guided by inter-molecular interactions often follow a path that decreases the internal energy of the binding proteins. Such mechanism increases the binding affinity. The energy decrease can be achieved by improving the interface packing upon binding. The prevalence of the disorder-to-order interface transitions over the reverse transitions corroborates this hypothesis (30). The two-sample t-test showed statistical significance of the difference between the SM-bands' centers at the 5% level for all proteins, except ubiquitin (Table S2). LM resulted in a significant decrease of the distance between the centers and a cancellation of the statistical significance of the difference between the centers of the unbound energies and bound energies for dimeric states of lysozyme C and ovomucoid, and multimeric states of PTI and lysozyme C. Comparison of the centers of the bound spectra of the structures extracted from dimers and multimers (Table S3) shows that LM resulted in statistically significant difference between the corresponding centers for RNase and PTI. For these proteins, the center of the dimeric bound band of the dimer states is lower than the center of the multimeric bound band.

An overlap between the unbound and bound energies, shown in Figure 1, suggests that conformational selection of a bound conformation may be guided by entropy. Indeed, if two protein states have equal energies $E_1 \approx E_2$, then the choice of the bound state is guided by the entropy contribution to the binding free energy $S_{1,2} = R ln N_{1,2}$ (36), where $N_{1,2}$ is the number of microstates associated with the unbound macrostates 1 and 2, and $R$ is the gas constant. An unbound macrostate contributes $-(E-TS) = -E + RlnN$ to the binding free energy. Therefore, a less-populated state increases the negative value of the binding free energy to a lesser extent. Thus, if $E_1 \approx E_2$, a less-populated conformation is the most effective binder, in agreement with the concept of conformational selection (37). Lower entropy means that protein conformations that are candidates for its bound state reside in narrow energy basins formed by the intra-molecular interactions. We intend to verify this hypothesis in our future study. The entropy-driven modeling of the bound-like conformations may improve performance of protein-protein docking algorithms, which commonly neglect entropy effects.

*Energy spacing, fluctuations and conformational changes*

Figure 3 summarizes calculations of the energy spacing in the ensembles. The spacing in the unbound ensembles averaged over the protein set is 2.9 kcal/mol, which is two times less than the average spacing in the bound ensembles. On average, the dimeric and multimeric states are separated by 6.0 kcal/mol and 6.4 kcal/mol accordingly. Larger variations of the energy spacing in bound ensembles may be a result of a smaller size of these ensembles (Table 1). Considering the proteins separately, one can see that the energy spacing between the unbound states varies between 0.8 and 4.6 kcal/mol, which includes the Hyeon and Thirumalai's estimate of 0-3kcal/mol for the energy landscape roughness or the barrier (38) and 3.2-3.5kcal/mol barrier measured by single-molecule dynamic force spectroscopy for a complex of GTPase Ran and the nuclear transport receptor importin-b (39). Note that the average spacing of 2.9 kcal/mol is approximately equal to the maximum barrier found in the Hyeon and Thirumalai study (38) but less then the maximum barrier found by Nevo *et al* (39). The energy spacing between the bound states is larger and falls in the intervals of 3.5-10.5kcal/mol and 3.8-10.2kcal/mol for the dimeric and multimeric states correspondingly. The energy gap between the two lowest energy minima in the joint ensemble of the bound and unbound states is 0.9 kcal/mol for RNase A, 6.1 kcal/mol for PTI, 12.1 kcal/mol for CheY, 7.9 kcal/mol for ubiquitin, 6.5 kcal/mol for ovomucoid and 9.9 kcal/mol for Lysozyme C. The smallest energy gap was found for RNase A, which has the smallest interface and all-atom RMSD between bound and unbound states (Table 2). On the other hand, the largest gap in the CheY spectrum corresponds to the largest RMSD between its bound and unbound ensembles.

To explore whether an inter-state transition in principle can result from protein vibrations, one can estimate a number of the normal modes that needs to be involved in the protein energy fluctuation equal to the inter-state barrier. The molecular energy fluctuation in a canonical ensemble can be calculated as $\delta E = \sqrt{\langle \Delta E^2 \rangle} = RT\sqrt{c_V}$, where $c_V = c_T + c_R + c_o(T)$ is a specific heat capacity per molecule at constant volume, $c_T = c_R = 3/2$ are the contributions of three translational and three rotational degrees of freedom, and $c_o(T) = \sum_i c_o^i(T)$ is the contribution of the $3N_{at} - 6$ internal vibrations in a protein with $N_{at}$ atoms. For low-energy modes $\hbar\omega_i \ll kT$ and therefore $c_o^i(T) \approx 1$ and $c_o(T) \approx n + \sum_{i=n+1} c_o^i(T)$, where $n$ is a number of the low-energy modes. For high-energy modes $\hbar\omega_i \gg kT$ and $c_o^i(T)$ exponentially goes to zero. One can find the low-bound estimate for the number of low-energy modes needed for a transition between any two states separated by the $\delta E$ barrier: $n = (\delta E/RT)^2 - 3$. It gives 23 and 32 normal modes for the standard ambient temperature of 298.15K and the barrier of 3.0 and 3.5 kcal/mol correspondingly. Since functional modes are often found among the lowest 20-30 modes (40-44), we can suggest that the protein vibrations indeed can excite the inter-state transitions related to protein function. Thus, an external stimulus (*e.g.*, a ligand or a partner-protein) may not be needed for changing protein conformation, which supports the conformational selection paradigm. Interestingly, many more modes are needed for an energy fluctuation covering the distance between the two lowest states in all the proteins in our set, except RNase A, which shows the smallest distance between bound and unbound ensembles. This is supported by a study of conformational changes in myosin, calmodulin, NtrC, and hemoglobin (45), which showed that the first 20 modes contribute ≤50% of the conformational changes in these molecules. The first 30 modes of the [AChE$_T$]$_4$–ColQ complex account for 75% of the conformational change in the tetramer (43). For a typical 1000 atoms protein having 2994 normal modes, a fluctuation of 12.1 kcal/mol is achieved when at least 13% of the normal modes get involved. It was shown that the protein energy fluctuation can increase up to 38 kcal/mol (46), which is more than enough for a transition over the largest gap/barrier considered in this study.

*Distance between unbound and bound conformational ensembles and binding mechanisms*

The size of the ensemble is controlled by ambient parameters (temperature, pH, salt concentration *etc*.) and dependent on protein sequence composition (Table 2). Protein-protein interactions can either select a group of bound-like conformations from the unbound ensemble or transform the whole ensemble into a new group of bound-like conformations. To find out which mechanism takes place, all-atom and interface RMSDs were calculated between all bound and unbound structures (Table 2). The smallest distance between the bound and unbound ensembles was found for RNase A and lysozyme C. RNase A shows the all-atom/interface RMSD of 0.7/0.3 Å. The ensembles of lysozyme C are separated by 0.8/0.7 Å of the all-atom/interface RMSD. Interestingly, the unbound ensemble of lysozyme C encompasses X-ray structures only, and RNase A has the second largest share (35%) of X-ray structures in its unbound ensemble among the proteins in our set. The low bound of the unbound-to-bound all-atom and interface RMSDs varies within 0.7-1.9 Å and 0.3-1.6 Å. The largest distance between the bound and unbound ensembles of CheY corresponds to the smallest overlap between their bound and unbound SM spectra (Figure 1), which disappears after LM. Thus, the majority of the SM and all LM bound conformations of CheY have lower energies than the unbound ones. It is likely that the entropy discussed above makes these lower-energy states unfavorable for the unbound ensemble.

In addition to the RMSD analysis, we calculated the share of the unbound residues within 1 Å of their bound conformations for all pairs of the bound and unbound structures (Figures 4 and 5). This metric also showed the lowest similarity between the CheY ensembles at 0.39 level (39% of all the residues, Fig. 5). The ensembles of RNase A, lysozyme C, and PTI had the highest similarity at 0.9, 0.89 and 0.86 levels correspondingly. Comparison of the bound and unbound interfaces revealed 13 unbound structures of RNase A with *all* interface residues within 1Å from the bound conformations (Fig. 4). Note that 2 Å is a typical size of a rotamer (32). Thus, dimerization of RNase A with another protein can be completely described by the conformational selection mechanism (5-8). Contrary to that, forming a multimer involving RNase A invokes induced fit to expose its C-terminal (Fig. S2), which forms an interface β-strand that swaps with the N-terminal helix in the RNase trimer. None of the unbound structures has the exposed C-terminal. This further suggests that some proteins may employ various binding mechanisms, from the induced fit, to "lock-and-key" and conformational selection, and their combination, depending on the binding partner (6, 7).

Interestingly, flash cooling used to determine approximately 90% of macromolecular structures (34) results in a 0.2 - 0.8 Å backbone RMSD between the structures determined at cryogenic and room temperatures (35). It can also change the conformational distribution of up to one third of the protein side chains (47). Taking this into account, we can assume that crystallographic conditions may distort the structure by 1 Å RMSD of *all* atoms. The low bound of the *all-atom* RMSD between bound and unbound ensembles (Table 2) suggests that, in addition to RNase A, the conformational selection likely guides binding processes of pancreatic trypsin inhibitor, ubiquitin (11), and lysozyme C.

## Materials and Methods
### Generation of the Protein Set

To compile a set of proteins with multiple bound and unbound conformations, a subset of protein complexes with small changes in the backbone upon binding (all-atom RMSD ≤ 2 Å) was selected from the non-redundant DOCKGROUND set 3.0 (48). The subset covers 71% of the DOCKGROUND set of 233 complexes. The subset was narrowed down to proteins that are monomers in the unbound state of the biological assembly. Their sequences were used to identify homologous proteins in PDB (sequence identity > 98% by BLAST (49)). The unbound protein structures with small ligands were excluded. All PDB entries found for each query-protein were put into three ensembles: unbound monomers, dimers and multimers. Only proteins with more than five unbound and bound structures were retained. Selected structures were analyzed for disordered residues and mutations. If some of the structures had a disordered terminal, it was deleted in all members of the ensemble. All fragments with ≤ 3 disordered residues at the interface and ≤ 5 at the non-interface were reconstructed by a program Profix from the Jackal package (http://wiki.c2b2.columbia.edu/honiglab_public/index.php/Software). Structures with disordered fragments longer than five residues were discarded. Point mutations were reversed by Profix. The resulting set consisted of six proteins (Tables 1 and S1) with multiple X-ray and NMR-derived bound and unbound conformational states and 100% sequence identity between the states.

### Minimization Protocol

The MMTSB Tool Set (50) and the Generalized Born method that calculates Born radii by analytic volume integration (CHARMM: GBMV method 2) were used to minimize solvation free energy of the proteins (51, 52). The method was parameterized to accurately reproduce electrostatic solvation energies from standard Poisson theory. A non-polar contribution to the solvation free energy was calculated by the ASP model considering the exposed surface area (53). Each protein was subjected to 50 steps of the steepest descent minimization (short minimization; SM) followed by $10^4$ steps of the Adopted Basis Newton-Raphson minimization (long minimization; LM). The CHARMM22 force field was used. The dielectric constant was set to 1 for protein and 80 for solvent. In minimizations, a switching function was used for truncating long-range interactions between 16Å and 18Å. Each bound protein was minimized within its complex to keep interface unchanged. The analysis showed that protein energy changed ≤1.5% between 500 and $10^4$ steps of LM. The average RMSD between all heavy atoms of the initial and minimized structures after short minimization was 0.1 Å. LM produced the average all-atom RMSD at 0.7 Å between the initial and the minimized structures. As can be seen from Table 2, LM did not change substantially the RMSD-based size of the conformational ensembles and the distance between the unbound and bound ensembles.

## Characterization of the energy spectrum

The ratio of the spectrum width in the ensemble of the SM and LM unbound structures to the lowest energy was calculated as the absolute value of $100\% \cdot (\Delta E_1 + \Delta E_2)/E_L$, where $E_L$ is the lowest protein energy in the joint ensemble of bound and unbound structures, and $\Delta E_{1,2}$ are the energy span in the unbound ensemble after the SM and LM correspondingly. If the energy spans overlap in the SM and LM ensembles, then the ratio was calculated as $100\% \cdot (E_{min} - E_{max})/E_L$, where $E_{min,max}$ are the lowest and the highest energies in the unbound spectrum.

The ruggedness of the energy landscape was calculated as $\bar{E}_{SM} - \bar{E}_{LM}$, where $\bar{E}_{SM}, \bar{E}_{LM}$ are the average energies in a protein ensemble after SM and LM accordingly. The energy spacing was calculated as the average distance between energy levels: $\sum_{i=1}^{N}(E_i^{LM} - E_{i+1}^{LM})/(N-1) = (E_H^{LM} - E_L^{LM})/(N-1)$, where $\{E_i^{LM}\}$ is an ordered set of the LM energies, $N$ is the number of structures in the LM ensemble.

### Acknowledgements
This study was supported by grant R01GM074255 from the NIH.

**Table 1.** Ensembles of bound and unbound proteins.

| Protein | Unbound structures | Bound structures [a] | |
|---|---|---|---|
| | | Dimers | Multimers |
| RNase A | 49 | 32 | 3 |
| Pancreatic trypsin inhibitor (PTI) | 27 | 18 | 29 |
| Chemotaxis protein CheY | 73 | 6 | |
| Ubiquitin | 394 | 8 | |
| Ovomucoid | 124 | 6 | |
| Lysozyme C | 45 | 19 | 24 |

[a]Considered separately from the other subunit(s) in the dimers/multimers.

**Table 2. Bound-to-bound and bound-to-unbound RMSDs.**

| Protein | | RMSD between bound structures, Å | | | | | | RMSD between bound and unbound structures, Å | | | | | |
|---|---|---|---|---|---|---|---|---|---|---|---|---|---|
| | | All atoms | | | Interface | | | All atoms | | | Interface | | |
| | | I[a] | SM[b] | LM[c] | I | SM | LM | I | SM | LM | I | SM | LM |
| **RNase A** | min | 0.3 | 0.3 | 0.3 | 0.1 | 0.1 | 0.2 | 0.7 | 0.8 | 0.7 | 0.3 | 0.3 | 0.4 |
| | max[d] | 7.1 | 7.2 | 7.2 | 11.6 | 11.8 | 11.7 | 7.3 | 7.4 | 7.4 | 11.7 | 11.9 | 11.8 |
| **PTI** | min | 0.1 | 0.1 | 0.3 | 0.0 | 0.0 | 0.2 | 1.0 | 1.0 | 0.9 | 1.1 | 1.1 | 1.0 |
| | max | 2.3 | 2.4 | 2.3 | 2.7 | 2.7 | 2.7 | 2.5 | 2.5 | 2.4 | 2.7 | 2.7 | 2.8 |
| **CheY** | min | 0.6 | 0.6 | 0.6 | 0.5 | 0.5 | 0.6 | 1.9 | 1.9 | 1.8 | 1.6 | 1.6 | 1.3 |
| | max | 1.2 | 1.1 | 1.2 | 1.7 | 1.7 | 1.8 | 2.9 | 2.9 | 3.0 | 3.1 | 3.1 | 2.9 |
| **Ubiquitin** | min | 0.3 | 0.1 | 0.4 | 0.4 | 0.1 | 0.5 | 1.0 | 1.0 | 0.9 | 1.1 | 1.1 | 1.0 |
| | max | 1.6 | 1.7 | 1.5 | 2.4 | 2.5 | 2.3 | 4.1 | 4.1 | 4.0 | 5.9 | 5.9 | 6.1 |
| **Ovomucoid** | min | 0.6 | 0.6 | 0.7 | 0.5 | 0.5 | 0.5 | 1.2 | 1.1 | 0.9 | 1.3 | 1.2 | 1.1 |
| | max | 1.2 | 1.2 | 1.2 | 1.6 | 1.6 | 1.8 | 2.2 | 2.2 | 2.1 | 3.5 | 3.4 | 3.1 |
| **Lysozyme C** | min | 0.2 | 0.2 | 0.4 | 0.1 | 0.1 | 0.2 | 0.8 | 0.8 | 0.8 | 0.7 | 0.7 | 0.7 |
| | max | 1.8 | 1.8 | 1.9 | 3.0 | 3.2 | 3.1 | 1.9 | 1.9 | 1.9 | 3.1 | 3.1 | 3.3 |

[a]Initial (not minimized) protein structures.
[b]Structures subjected to 50 steps of the steepest descent minimization (short minimization).
[c]Structures subjected to 50 steps of the steepest descent minimization, followed by $10^4$ steps of the Adopted Basis Newton-Raphson minimization (long minimization).
[d]Minimum and maximum RMSDs.

**Captions to the figures**

1. Energy spectrum of the unbound and bound proteins. (A) PTI, (B) RNase A, (C) Lysozyme C, (D) CheY, (E) Ovomucoid and (F) Ubiquitin. SM and LM indicate the spectrum after the short minimization (50 steps of the steepest descent minimization) and long minimization (SM followed by $10^4$ steps of the Adopted Basis Newton-Raphson minimization). U, D and M are unbound proteins (green), proteins crystallized as dimers (red), and multimers (blue). Open squares with error bars show the average energy (the band's center) and the standard deviation.

2. The ratio of the ensemble width after long minimization to the lowest energy in the joint ensemble. The data is shown for unbound (■) and bound ensembles, extracted from dimers (●) and multimers (○).

3. Energy spacing at the bottom of the folding funnel. The figure shows the energy gap between the two lowest energy minima (▲) in the joint ensemble of the bound and unbound states and the average distance between energy minima after the long minimization in the unbound (■) and bound ensembles, extracted from dimers (●) and multimers (○).

4. Similarity of bound and unbound interface conformations. (A) PTI, (B) RNase A, (C) Lysozyme C, (D) CheY, (E) Ovomucoid and (F) Ubiquitin. The similarity is calculated for each pair of bound and unbound structures as the share of the unbound interface residues within 1 Å RMSD from the bound interface residues. Bars and circles show the interface similarity between dimeric/multimeric and unbound conformations accordingly. The horizontal axis shows conformation in the bound ensembles.

5. Similarity of bound and unbound structures. (A) PTI, (B) RNase A, (C) Lysozyme C, (D) CheY, (E) Ovomucoid and (F) Ubiquitin. The similarity is calculated for each pair of bound and unbound structures as the share of the unbound residues within 1 Å RMSD from the bound ones. Bars and circles show the similarity between dimeric/multimeric and unbound conformations accordingly. The horizontal axis shows conformation in the bound ensembles.

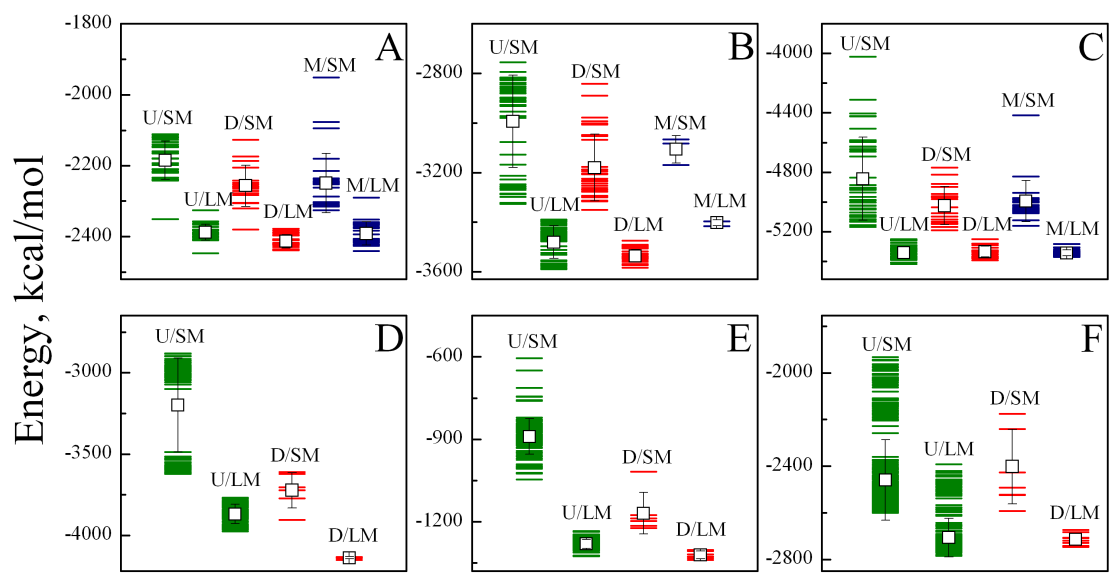

**Figure 1**

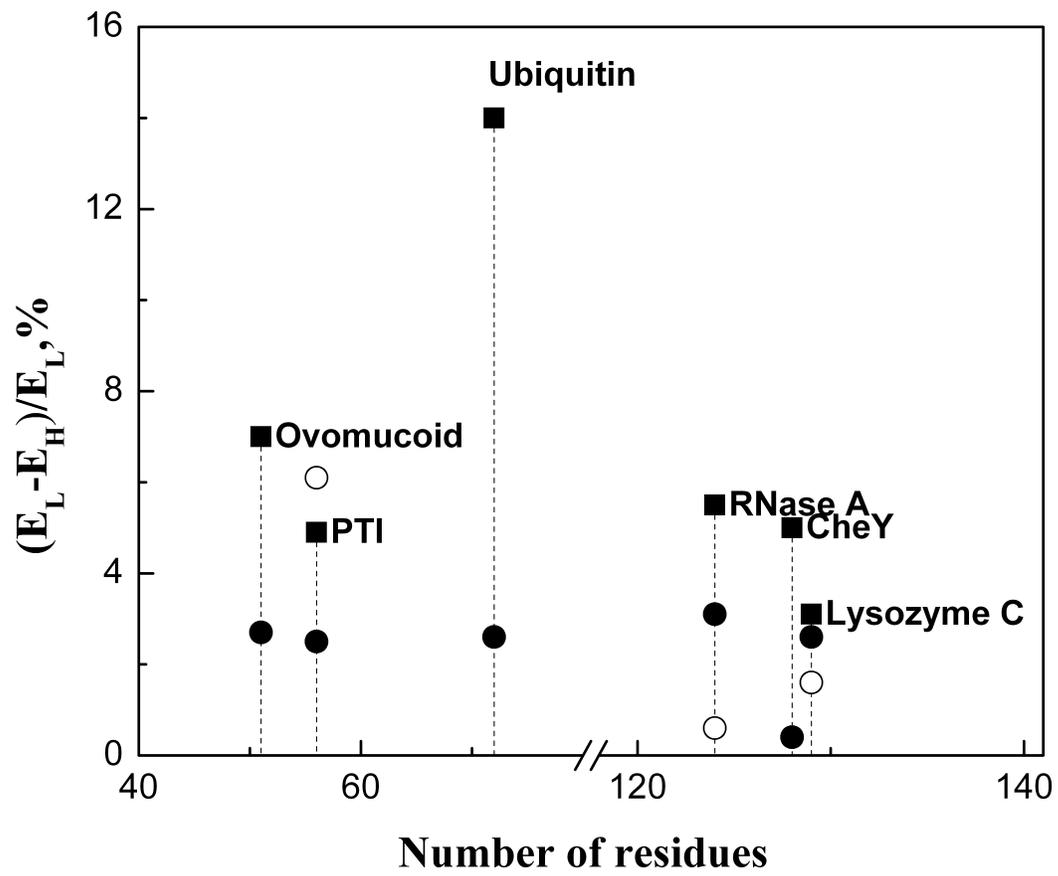

Figure 2

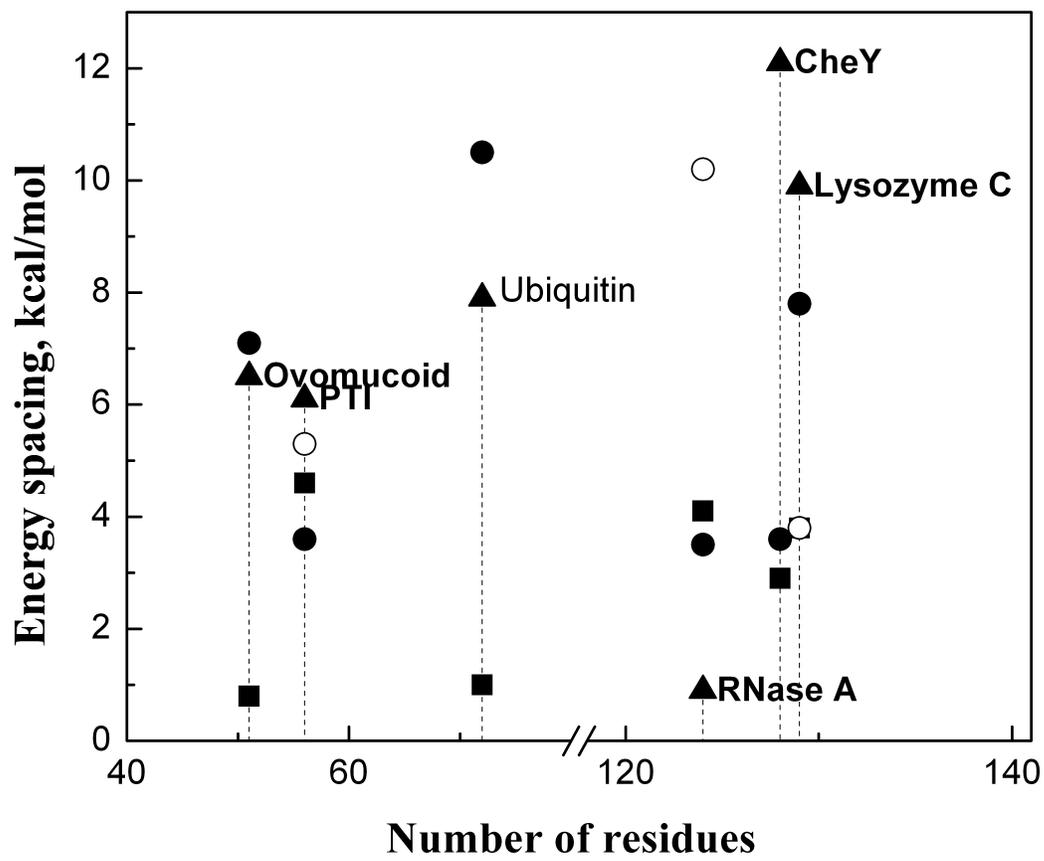

**Figure 3**

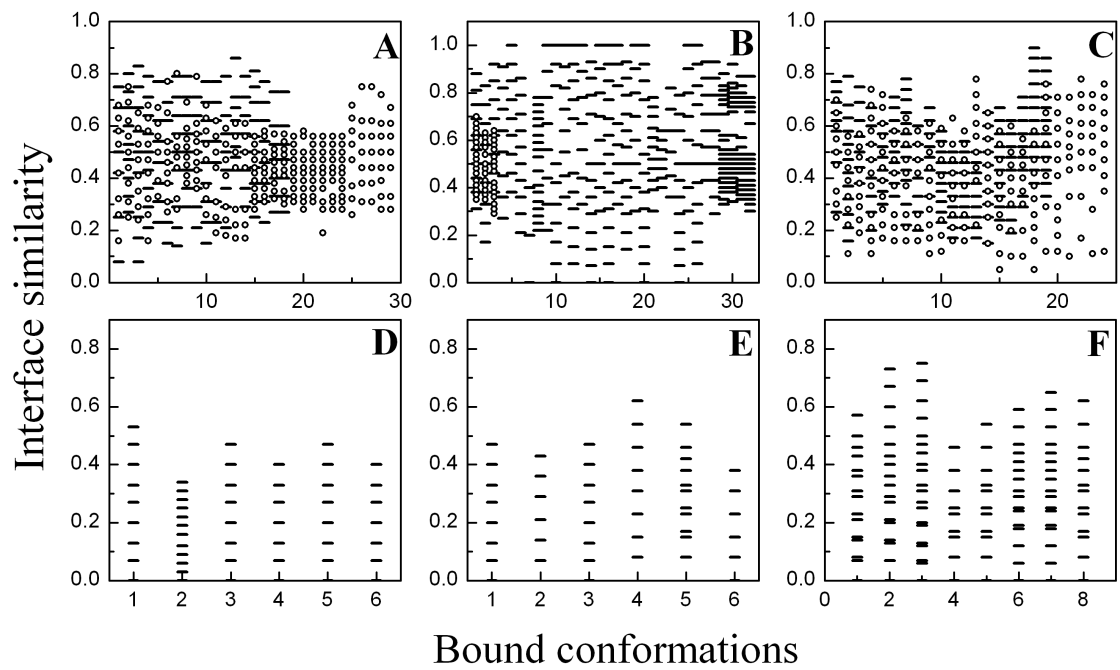

**Figure 4**

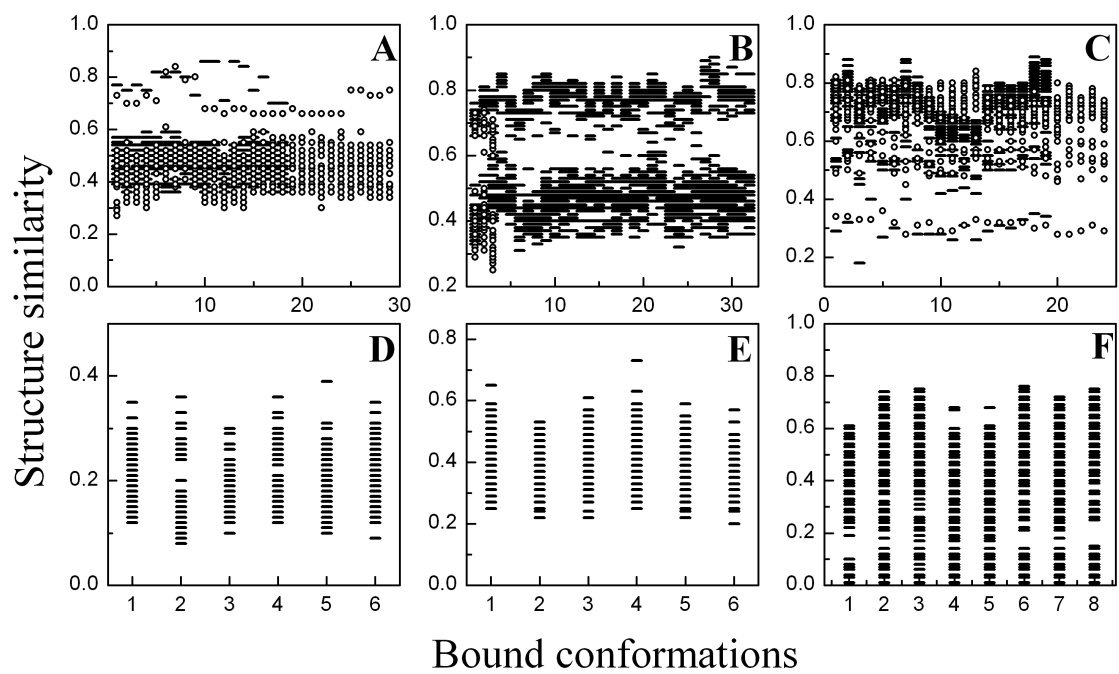

**Figure 5**

# SUPPORTING MATERIAL:
# Ensemble-based characterization of unbound and bound states on protein energy landscape

A.M. Ruvinsky, T. Kirys, A.V. Tuzikov, and I.A. Vakser

Table S1. PDB IDs of bound and unbound protein structures.

| Protein | Unbound structures | Bound structures [a] | |
|---|---|---|---|
| | | Dimers | Multimers |
| RNase | 9rat_A,[b] 8rat_A, 7rat_A, 6rat_A, 5rat_A, 4rat_A, 3rat_A, 2rat_A, 2aas_A(N,32)[c], 1rtb_A, 1rhb_A, 1rha_A, 1rbx_A, 1rat_A, 1kf8_A, 1kf7_A, 1kf5_A, 1fsa_A | 1bzq_A,B,C,D, 2p45_A, 2p49_A,2p48_A,2p47_A, 2p44_A,2p43_A, 2e33_B, 1dfj_E, 3ev6_A,B, 3ev5_A,B, 3ev4_A,B, 3ev3_A,B, 3ev2_A,B, 3ev1_A,B, 9rsa_A,B, 3jw1_A,B, 3euz_A,B, 2p4a_A,C | 1js0_A,B,C |
| Pancreatic trypsin inhibitor (PTI) | 4pti_A, 1pit_A(N,20), 1oa6_5(N,3), 1oa5_5(N,3) | 3tgk_I, 3tgj_I, 3tgi_I, 1f5r_I, 3tpi_I, 3fp8_I, 3fp6_I, 3btk_I, 2tpi_I, 2tgp_I, 2ptc_I, 2kai_I, 2ijo_I, 1tpa_I, 1fy8_I, 1bzx_I, 3gym_I,J | 2hex_A,B,C,D,E, 1mtn_H,D, 1cbw_I,D, 1bz5_A,B,C,D,E, 1bhc_A,B,C,D,E,F,G,H,I,J, 1b0c_A,B,C,D,E |
| Chemotaxis protein CheY | 1djm_A(N,27), 1cey_A(N,46) | 1ffg_A, 1kmi_Y, 1a0o_A,C,E,G | |
| Ubiquitin | 1ubq_A, 1ubi_A, 2nr2_A(N,144), 1xqq_A(N,128),2kn5_A(N,50), 2klg_A(N,20),2jzz_A(N,20), 1v81_A(N,10),1v80_A(N,10), 1d3z_A(N,10) | 1p3q_V, 1s1q_B,D, 1yd8_U,V, 2c7m_B, 2fid_A, 1wrd_B | |
| Ovomucoid | 1tus_A(N,12), 1tur_A(N,12), 1omu_A(N,50), 1omt_A(N,50) | 1r0r_I, 3sgb_I, 1ppf_I, 1cho_I, 1sgr_I, 1ds2_I | |
| Lysozyme C | 8lyz_A, 7lyz_A, 6lyz_A,6lyt_A,5lyz_A,5lyt_A,4lyz_A,4lyo_A,4lym_A,3lyz_A, 3lym_A,3exd_A,2zq4_A,2yvb_A,2lyz_A,2lym_A,2hso_A,2hs9_A,2hs7_A,2epe_A, 2cds_A, 2c8p_A, 2c8o_A,2aub_A, 2a6u_A, 1xek_A,1xej_A, 1xei_A, 1ved_A, 1vdt_A, 1vds_A,1vdq_A, 1uig_A, 1rfp_A, 1lzt_A, 1lza_A, 1lyz_A, 1lyo_A,1lsf_A,1lse_A, 1lsd_A,1lsc_A,1lsb_A,1lsa_A, 1jpo_A | 1uuz_D, 1zvy_B, 1zvh_L, 1zv5_L, 1xfp_L, 1sq2_L, 1rjc_B, 1ri8_B, 1jtt_L, 3g3a_B,D,F,H, 2znx_Z,Y, 2znw_Z,Y, 2i25_M,L | 3hfm_Y, 3d9a_C, 2yss_C, 2eks_C, 2eiz_C, 2dqj_Y, 2dqi_Y, 2dqh_Y, 2dqg_Y, 2dqe_Y, 2dqd_Y, 2dqc_Y, 1yqv_Y, 1xgu_C, 1xgt_C, 1xgr_C, 1xgq_C, 1xgp_C, 1vfb_C, 1ua6_Y, 1ndm_C, 1kir_C, 1kiq_C, 1kip_C |

[a] Considered separately from the other subunit(s) in the dimers/multimers.
[b] 9rat_A is an X-ray structure of a chain A from 9rat.
[c] 2aas_A (N,32) are 32 NMR structures of a chain A from 2aas.

Table S2. Two-sample t-test for equal means of energies of the unbound and bound ensembles.

| | P-values, % | | | |
| --- | --- | --- | --- | --- |
| | Dimeric ensemble *vs* Unbound ensemble | | Multimeric ensemble *vs* Unbound ensemble | |
| Protein | SM[a] | LM[b] | SM | LM |
| RNase A | 0.0001 | 0.0002 | 4 | 0.2 |
| CheY | 0.00006 | $8.8*10^{-41}$ | - | - |
| PTI | 0.02 | 0.13 | 0.11 | 44 |
| Ubiquitin | 35 | 42 | - | - |
| Ovomucoid | 0.02 | 41 | - | - |
| Lysozyme C | 0.08 | 70 | 0.4 | 58 |

Table S3. Two-sample t-test for equal means of energies of the bound structures extracted from dimers and multimers.

| | P-values, % | |
| --- | --- | --- |
| | Multimeric ensemble *vs* Dimeric ensemble | |
| Protein | SM | LM |
| RNase A | 13 | 0.1 |
| PTI | 73 | 2 |
| Lysozyme C | 47 | 96 |

[a]Short minimization.
[b]Long minimization.

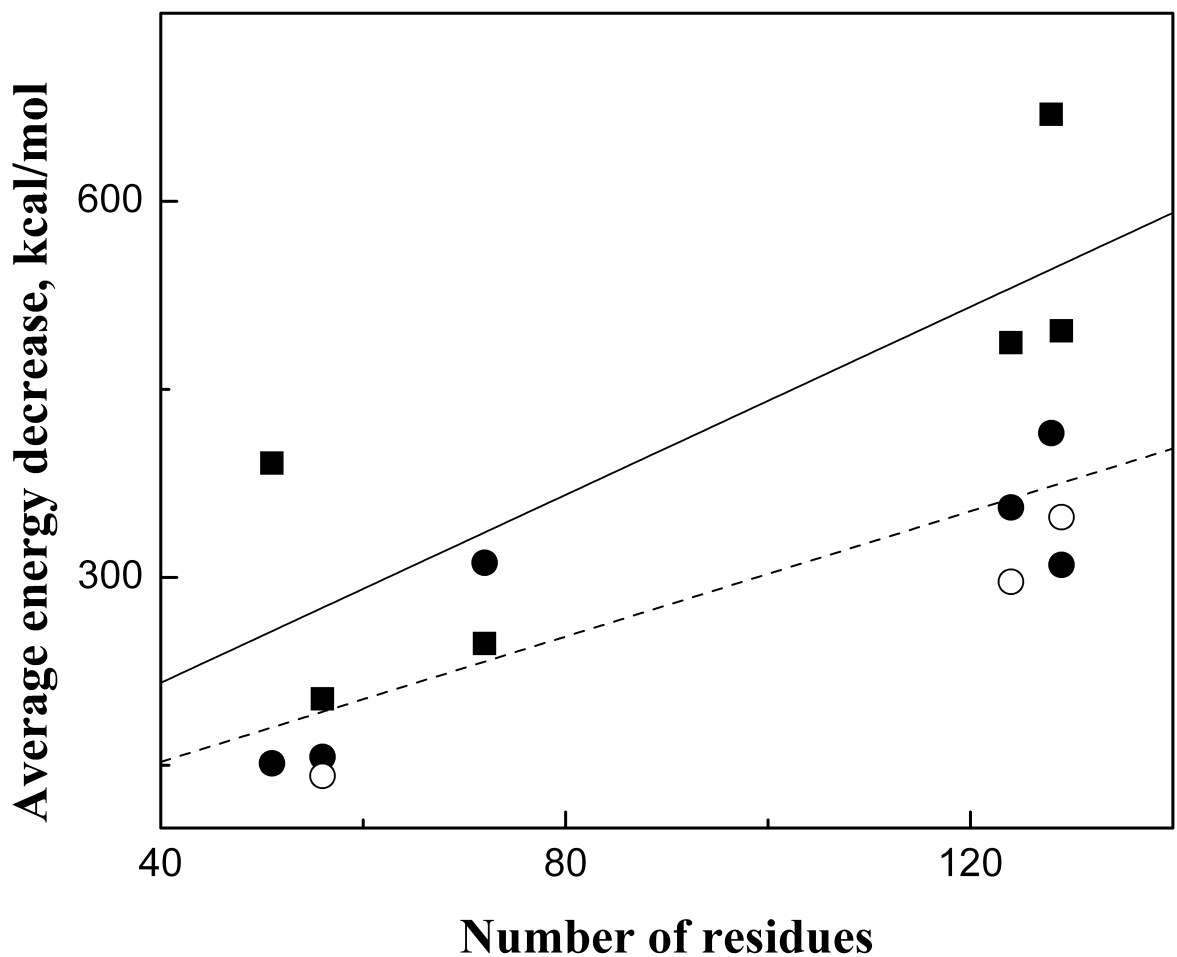

**Figure S1**. Ruggedness of the energy landscape. The ruggedness was calculated as a difference between the average energies of a protein after short and long minimizations for each unbound (■) and bound ensembles, extracted from dimers (●) and multimers (○). The solid and dashed lines show linear fit to the data for the unbound proteins and the ones co-crystallized as dimers. The slopes of the solid and dash lines are 3.8 and 2.5 kcal/mol per residue.

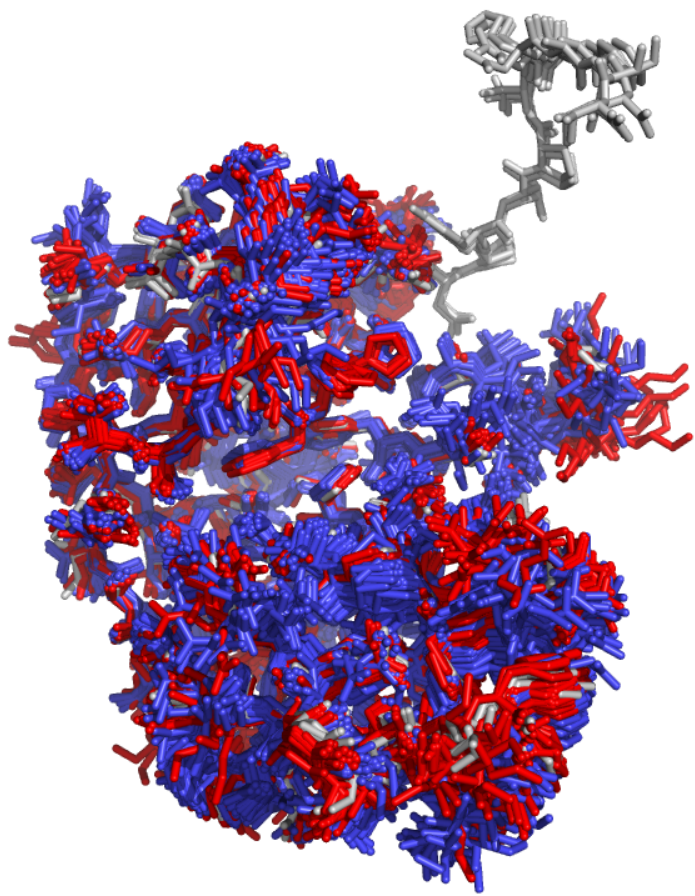

**Figure S2**. Structure ensemble of RNase A. The unbound, dimeric and multimeric structures are in blue, red and gray.